\documentclass{article}

\usepackage{arxiv}

\usepackage[utf8]{inputenc} % allow utf-8 input
\usepackage[T1]{fontenc}    % use 8-bit T1 fonts
\usepackage{hyperref}       % hyperlinks
\usepackage{url}            % simple URL typesetting
\usepackage{booktabs}       % professional-quality tables
\usepackage{amsfonts}       % blackboard math symbols
\usepackage{nicefrac}       % compact symbols for 1/2, etc.
\usepackage{microtype}      % microtypography
\usepackage{lipsum}		% Can be removed after putting your text content
\usepackage{graphicx}
\usepackage[square,numbers]{natbib}
\usepackage{doi}

\title{Under the Skin of Foundation NFT Auctions}

\author{
	MohammadAmin Fazli \\
	Department of Computer Engineering\\
	Sharif University of Technology\\
	Tehran, Iran \\
	\texttt{fazli@sharif.edu} \\
	\And
	Ali Owfi \\
	Department of Computer Engineering\\
	Sharif University of Technology\\
	Tehran, Iran \\
	\texttt{owfi.yojam@yahoo.com} \\
	\And
	Mohammad Reza Taesiri \\
	Department of Electrical and Computer Engineering\\
	University of Alberta\\
	Edmonton, Canada \\
	\texttt{taesiri@ualberta.ca} \\
}

\hypersetup{
pdftitle={Under the Skin of Foundation NFT Auctions},
pdfsubject={q-bio.NC, q-bio.QM},
pdfauthor={Mohammad Reza Taesiri, Ali Owfi, MohammadAmin Fazli},
pdfkeywords={Non Fungible Token, NFT, Blockchain, Digital Asset, Auction, Fraudulent Activities},
}

\begin{document}
\maketitle

\begin{abstract}
Non Fungible Tokens (NFTs) have gained a solid foothold within the crypto community, and substantial amounts of money have been allocated to their trades. In this paper, we studied one of the most prominent marketplaces dedicated to NFT auctions and trades, Foundation. We analyzed the activities on Foundation and identified several intriguing underlying dynamics that occur on this platform. Moreover, We performed social network analysis on a graph that we had created based on transferred NFTs on Foundation, and then described the characteristics of this graph. Lastly, We built a neural network-based similarity model for retrieving and clustering similar NFTs. We also showed that for most NFTs, their performances in auctions were comparable with the auction performance of other NFTs in their cluster.
\end{abstract}

% keywords can be removed

\section{Introduction}\label{s:intro}
% \section{Introduction}

\subsection{Non Fungible Token and its history}

An NFT, which stands for Non-Fungible Token, is a digital asset stored on a blockchain that attests ownership over a digital item such as art, music, or a game. Although NFTs and cryptocurrencies have some similarities as they are both based on blockchain, and some of their functionalities are the same, NFTs have a characteristic that separates them from cryptocurrencies: their non-fungibility. That means each NFT is a unique item that does not have an equal and thus can not be traded equally with another NFT. In contrast, a bitcoin or a dollar bill is the same as any other bitcoin or a dollar bill, and a bitcoin or a dollar bill can be changed with any other bitcoin or a dollar bill. Furthermore, an NFT is different from a copyright, while an NFT certifies ownership over a file, it does not prohibit other people's access to copies of the original file, and so everyone can obtain a copy of the original, only the ownership over the original file is tracked via blockchain.  

Arguably, the idea of non-fungible tokens can be traced back to 2012, when Colored Coins were introduced \cite{rosenfeld2012overview}. Colored Coins are tokens that represent real-world objects and assets and certify their ownerships. Although these tokens were a novelty and opened new windows of opportunity for the future of blockchain technology, they did not face the same success as their future successor NFTs. One reason was that these tokens were implemented on the bitcoin blockchain, which could not provide a ground that could capture all the intricacies that non-fungible tokens required as the implementation of NFTs needed a more malleable blockchain\cite{Steinwoldmedium}. 

In 2017, as Ethereum \cite{wood2014ethereum} rose to prominence, it provided a better foundation for the idea of unique interchangeable tokens that were set off a few years back, and so the concept behind NFTs received a better opportunity to grow and develop this time. Although there already were games that utilized blockchain for issuing their in-game asset, such as Spells of Genesis\cite{SpellsofGenesis}, many game-like and collectible projects that utilized blockchain technology by using NFTs in their games rose in 2017. Cryptopunks \cite{CryptoPunks} created and released 10000 unique pixel art characters for free at the time, which were swiftly collected, and from that time forth, they have been thriving in secondary markets, being bought and sold again. Up to this date, the overall sales from these 10000 NFTs have gone over 1.2 Billion Dollars. Cryptokitties, \cite{Cryptokitties}, a gamified NFT project that enabled players to trade and breed unique virtual cats, was another NFT project that bolstered forward NFTs into the mainstream due to its substantial success and popularity. 

Very soon, NFTs picked up steam in the following years, and a trend started around it. The public became more aware of the existence of NFTs, and the opportunities they could offer. Many platforms for trading NFTs began to be established, and many people started to mint NFTs on blockchains and traded them. The trend went on for a few years until, in the first months of 2021, the world of NFTs finally exploded, and many NFTs were sold for unforeseen prices. Even though NFTs have now made a prominent presence within the crypto community, it is believed that NFTs are still in their early stages as many aspects of NFTs can be further developed, and the possibilities that they could grant us have to be explored in the following years to come.

\subsection{NFT auctions and Foundation website}

After NFTs became an accepted digital asset in the popular view and after the wave of attention that NFTs have received in the past few years, many platforms for trading these tokens were established, such as Rarible, Mintable, Nifty Gateway. Foundation \cite{Foundation}, created in February of 2021, is another NFT marketplace that enables NFT creators to sell their artworks and other digital assets via online auctions based on Ethereum. This website is younger than many of its competitors, but it has quickly become one of the most popular of these platforms. Foundation has some characteristics and rules that separate this website and the behavior of its community of NFT traders from many other NFT marketplace websites. Perhaps, the principal property of this website is the fact that it has a closed society, meaning that even though anyone with an Ethereum wallet can bid and buy NFTs, only those who have been invited or have gone through a waiting queue can mint NFTs and become creators on this website. Furthermore, the process of how people can get invited to become creators on Foundation and how these invitation letters are distributed among the existing creators are pretty intriguing as the mechanism that Foundation has set for becoming a creator on this website results in some shady behaviors within their community, which are discussed in further sections of this paper. Another property that Foundation has is that, unlike some other NFT marketplace websites, every action requires a gas fee. That means we should assume that there is a strong incentive behind every activity on this website.

\subsubsection{Type of Actions that take place on Foundation}

There are several types of activities that can take place on Foundation, which are defined below.

\begin{itemize}
    \item Mint
    
    Minting is simply creating an NFT, and it is how digital art becomes an NFT by becoming part of the Ethereum blockchain. Just as a physical coin becomes valid after it becomes minted, digital art becomes a valid NFT certified by the Ethereum blockchain after it gets minted.
    
    \item List
    
    After an NFT becomes minted, the owner of the NFT can list that it for an auction with a minimum price which is referred to as reserve price. Then, the corresponding auction only starts after the first bid is received. The auction lasts for 24 hours, and it gets extended to 15 minutes whenever a bid is received in the last 15 minutes of the auction.
    
    \item Unlist
    
    To undo the effects of listing and bring an NFT to the state where no auction can be started, the owner of a listed NFT can unlist their NFT.

    \item Bid
    
    Whenever an NFT is listed for an auction with a determined reserve price, people interested in collecting that NFT can bid on it higher than the previous bids or the reserve price if there are no bids. 
    
    \item Settle
    
    When an auction ends, the winner becomes determined, but for the auction to be done entirely and for the effects of changing the ownership and transferring the money to take place, either the NFT's owner or the auction winner has to settle that auction.
    
    \item Transfer
    
    The owner of an NFT can relinquish their ownership over that NFT to another person by transferring that NFT. After moving, the owner of the transferred NFT changes.
    
\end{itemize}

\subsection{Methodology}
In this paper, we explored and analyzed the underlying dynamics of auctions on Foundation. Using a specialized crawler, We crawled almost 65 thousand auctions data, including artwork files, descriptions, creators, transaction information with their dates and times, from Foundation. After data collection, we analyzed different aspects of these auctions using social network analysis, exploratory data analysis, and neural networks. To be more precise, We investigated the amounts of activities on this website from its starting date, explored the potential of some speculative behaviors such as buying an NFT with the hopes of selling it at a higher price on Foundation, and analyzed the mechanism that enables people to mint NFTs on Foundation and the profit opportunity that it may bring for some. Furthermore, we examined the NFT transfers that took place on Foundation and analyzed their corresponding graph, and finally, we tried to answer whether there exists a relationship between the actual content of an NFT and its price using deep neural networks.

\section{Auction Properties} \label{s:method}
% \section{Foundation NFT Auction Properties}

\subsection{Activity on Foundation}
\begin{figure}[!ht]
    \centering
    \includegraphics[width=0.95\textwidth]{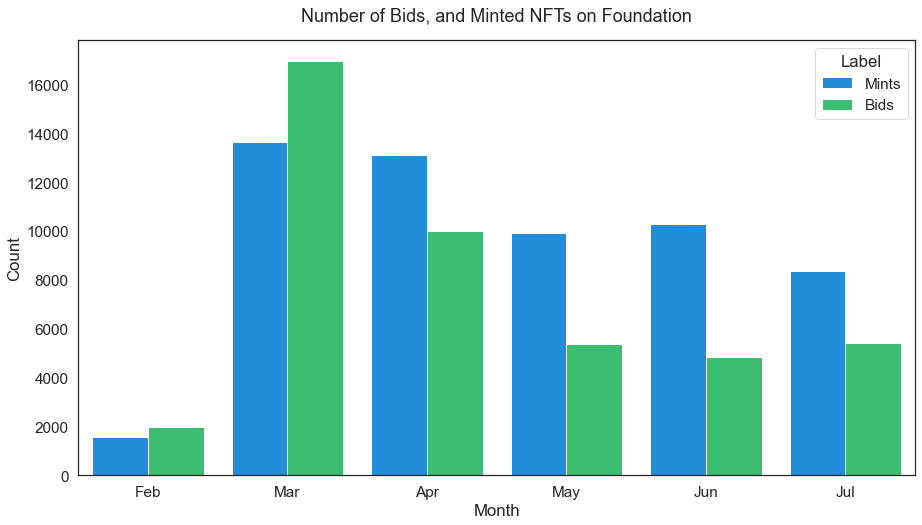}
    \caption{Number of bids in auctions and the number of NFTs minted on Foundation since the launch date of Foundation}
    \label{fig:bid_mint_histogram}
\end{figure}

The amount of activity on Foundation has not been constant since its start, and in fact, it has seen different levels of attention in different months of 2021. Figure \ref{fig:bid_mint_histogram} shows the number of bids and the number of minted NFTs on Foundation since its start. As this plot suggests, activity on Foundation reached its peak during March of 2021. It is worth mentioning that the record-breaking auction of ”Everydays — The First Five Thousand Days” by the artist Beeple which was settled for 69 million dollars and drew much attention to NFTs, also happened during the same period. Another interesting fact about figure \ref{fig:bid_mint_histogram} is that, while the number of bids has significantly dropped from the first months of 2021, the number of NFTs minted by creators has not seen a drastic decline. This contrast between minting and bidding activities could suggest that even though art collectors are now less inclined to buy NFTs on Foundation than in the first months, digital art creators have not lost hope in selling their artworks on Foundation.

\begin{figure}[!ht]
    \centering
\includegraphics[width=0.95\textwidth]{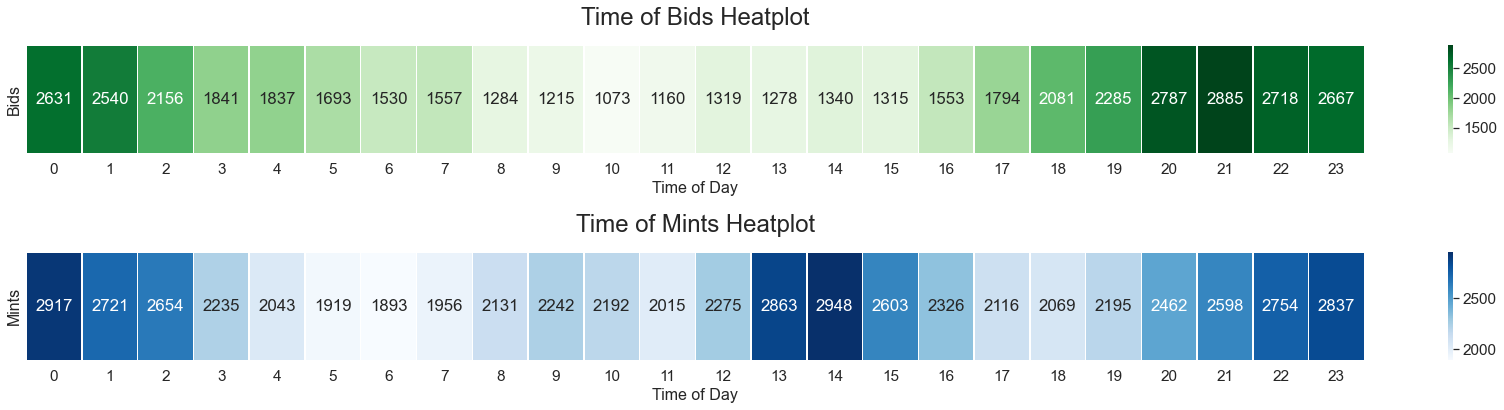}
    \caption{Daily minting and bidding activities of the Foundation community by hour}
    \label{fig:bid_mint_heatplot}
\end{figure}

Furthermore, figure \ref{fig:bid_mint_heatplot} indicates the busy hours in which most of the Bids and Mints happen. Even though both Minting and Biding activities show some relative similarities, such as their high chance of occurrence during 4 pm to 8 pm UTC, it is interesting that Minting activities also occur a lot at 9 am while that hour is among the least busy hours for bidding. This drastic contrast is intriguing and may suggest that NFT creators, who mint NFTs, are concentrated in a different time zone than the NFT collectors, who bid in NFT auctions.

\subsection{Second Hand Auctions}

The number of times that an NFT can be sold on an auction on Foundation is not limited, and this characteristic opens a possible window of opportunity for speculative behaviors among the collectors. A buyer might try and win an NFT in an auction, not for the personal value that they hold for that art, but in hopes of selling that NFT at a higher price. Thus, by its nature, an NFT auction marketplace website like Foundation has the potential to become a speculative trading hub, but the extent of this potential should be explored.

To further investigate the raised issue, we analyzed the dataset that we have crawled and reached some exciting statistics. Only 15.72 percent of the NFTs that have been successfully sold on their first auction have been listed for a second auction. That means collectors are not very eager to sell their collected NFTs for a second time, at least yet.
Moreover, while the first auction success rate - the chance that an NFT gets sold when it gets listed for its first auction - is 36.10 percent, the corresponding success rate for the second time auctions is only 3.97 percent which is a substantial drop. This significant decrease suggests the second hand NFT market on Foundation is not very active yet, but also, the price that the speculators list their collected NFTs for the second auction in comparison to the price that they have bought that NFT should be considered in determining why such a decrease of success rate between the second auctions and the first auctions exist. 

\begin{figure}[!ht]
    \centering
    \includegraphics[width=0.95\textwidth]{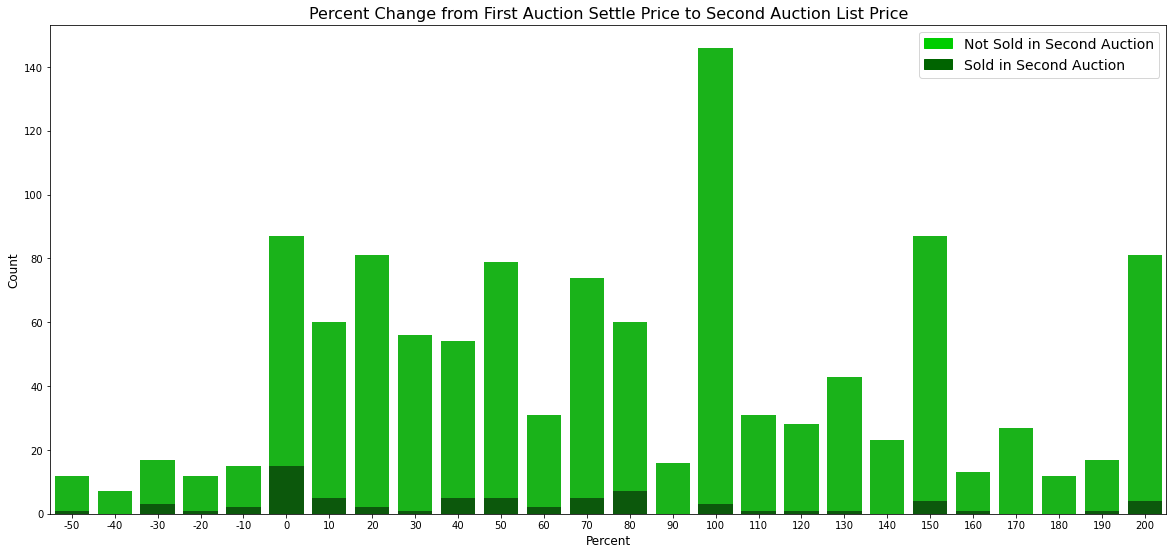}
    \caption{A histogram displaying percent change from first auction settle price to second auction list price.}
    \label{fig:second_list_to_settle_price}
\end{figure}

Figure \ref{fig:second_list_to_settle_price} depicts the price change of NFTs, from their first auction settle price to their second auction list price. In other words, it shows the target profit that those who wish to make money from buying an NFT and selling it again have. Different colors in this barplot also indicate that an NFT that was listed for a second auction was successfully sold or not.
There are some cases where the person who had bought an NFT and then had listed it for a second auction has made a significant profit -"A Kings Abomination" was sold for 0.666 ETH in its first auction but was resold for 50 ETH in its second auction \cite{AKingsAbomination}. However, in reality, as figure \ref{fig:second_list_to_settle_price} shows, most of the profits from reselling NFTs were not very significant, and that is not to mention that most of the NFTs are not being bought for a second time in the first place. It is also interesting to note that even some NFTs were listed and resold in their second auction for a lower price than they were bought in the first auction, which means the first auction winner has even sustained loss during the reselling process.

\subsection{Unlisting and Relisting}

\begin{figure}[!ht]
    \centering
    \includegraphics[width=0.95\textwidth]{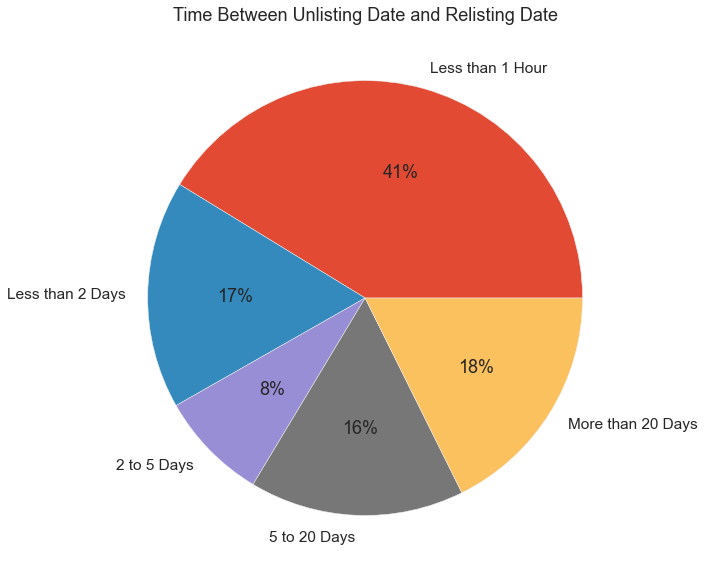}
    \caption{This pie chart depicts the time gap between unlistings and relistings of the NFTs that were unlisted and then listed again for a second time.}
    \label{fig:unlist_relist}
\end{figure}

Before an NFT can receive any bids and before its auction can be started on Foundation, the corresponding NFT needs to be listed with a defined reserve price by the owner. After an NFT is listed, if it gets any bids, a 24-hour auction begins, and the owner will not be able to undo the auction. An owner can also unlist an NFT they had listed previously, which reverses the effects of listing. In figure \ref{fig:unlist_relist}, we can observe that 41 percent of unlist-relists on Foundation have happened within 1 hour, and overall 64 percent of unlist-relists happen in just five days. While unlisting and NFT is quite a normal behavior, as an owner may not want to put their owned NFTs for an auction anymore, it is intriguing when someone unlists an NFT and then relists it after a short while, since it seems that this chain of actions accomplishes nothing. However, it should be noted that every action on Foundation costs money, so we should assume that there is an incentive behind every pattern on Foundation.

One explanation that could justify the unlist-relists that occur in a brief period is that NFTs that have been more recently listed, have a higher chance of being exhibited on the Foundation homepage. Another explanation for unlist-relists that happen in a longer time span is that the owner might want to unlist their NFT from Foundation so that they can list their NFT on another NFT marketplace platform with hopes of selling their NFT on the alternative marketplace, but when they can not sell it there, they relist their NFT on Foundation.

\subsection{Buying NFT to receive invitation}

Foundation is considered to be a closed community website in the sense that every person can collect other NFTs, but in order for someone to have the authority to mint an NFT and become a creator, that person either needs to receive an invitation from another creator, or get approved by Foundation, and this feature of Foundation is one of the main aspects of what distinguishes this website from its other counterparts such as Rarible, OpenSea, and Mintable. This feature has had a significant impact on Foundation in general and how the community on this website interacts with each other. For instance, it can be deduced that this characteristic of being a closed community has contributed a lot to the high NFTs selling success rate of Foundation in comparison to other NFT auction websites, as the concentration of verified real digital artists are more on this website due to its closed community nature. 

\begin{figure}[!ht]
    \centering
    \includegraphics[width=0.95\textwidth]{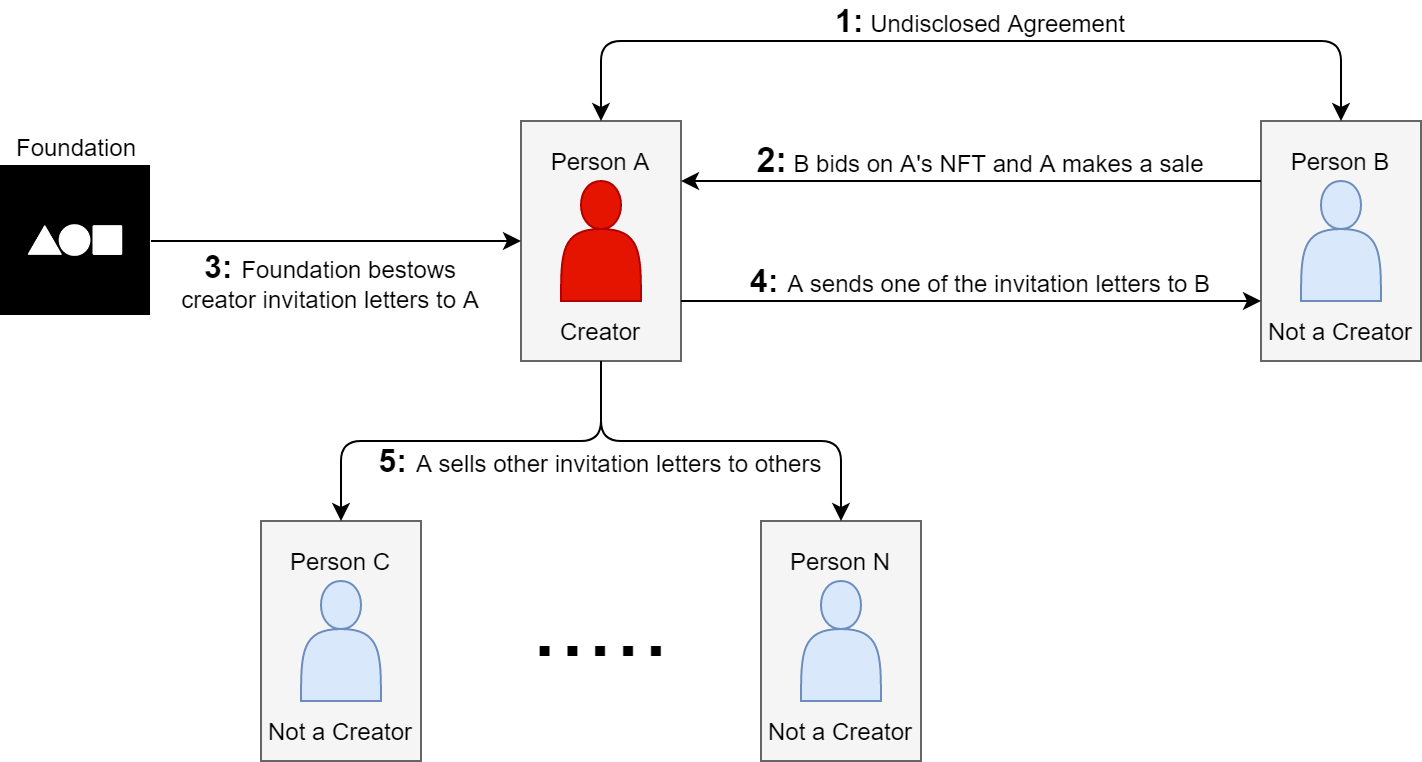}
    \caption{This diagram demonstrates a type of collusion that takes place on Foundation, where a person receives Foundation creator invitation letter by buying an NFT from a creator account.}
    \label{fig:buy_invitation_diagram}
\end{figure}

Moreover, as the popularity of NFTs and Foundation rose drastically in 2021, and as more digital artists found a new way to sell their arts, the demand for having the authority to be a creator on Foundation rose as well. However, since the supply of this creator authority on Foundation is limited and may take a long time to acquire with the waiting queue that Foundation has, it is very natural that some creators that have invitations will try to sell them to those who want a creator account. Furthermore, another mechanism on this site contributes to such a trade. A person who has received a creator account will be bestowed invitations to send to others only after they make a sale on Foundation, and this mechanism will further encourage those creators who still have not made a sale and thus have no invitations to send, to get involved in "buy NFT for invitation" trades. Figure \ref{fig:buy_invitation_diagram} depicts how such a trade works. In a "buy NFT for invitation trade", individuals A and B, where A has a creator authority on Foundation and B has not, reach an undisclosed agreement behind the website. Person B is promised to receive an invitation letter from person A if they bid on one of person A's listed NFTs, thus starting the auction for that NFT. 

\begin{figure}[!ht]
    \centering
    \includegraphics[width=0.95\textwidth]{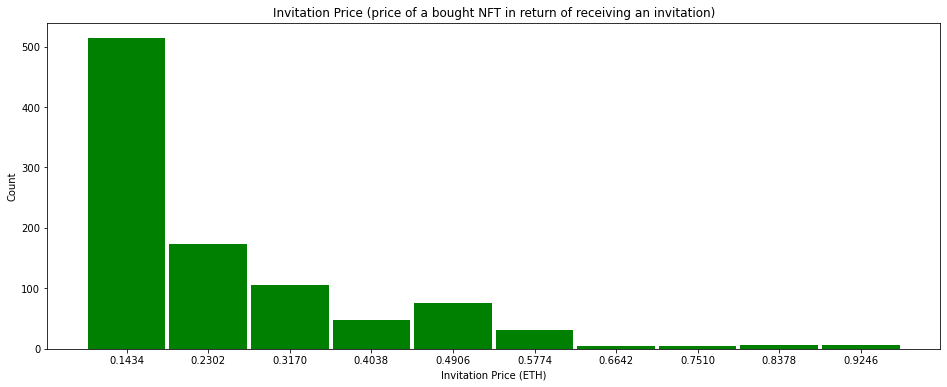}
    \caption{Prices of NFTs that were bought in exchange for receiving invitation letters on Foundation.}
    \label{fig:invite_price}
\end{figure}

We have analyzed the dataset that we have crawled in search of interactions between two accounts where one bought an NFT from a creator account and then received an invitation from that creator. Even though we can not make sure whether the two involved individuals in these types of interactions actually colluded with each other, the fact that someone was invited by a person they had bought an NFT from is quite intriguing and can be a strong indicator of collusion. Figure \ref{fig:invite_price} displays the price of the NFTs that were bought in the described type of interactions, and their price can be a good measurement for evaluating the value of having a creator account on Foundation. The average price of these NFTs was 0.3716 Ether, and we can assume this amount as an approximation of Foundation creator rights value.

\section{Transfer Networks} \label{s:tn}
% \section{Transfer Network}

On Foundation, there are several types of interactions that two accounts can have with each other such as bidding, settling, inviting, and transferring. These interactions can indicate the closeness of the two accounts, especially since every action on Foundation costs a gas fee. Still, naturally, some of these interactions display a stronger connection. Transferring is when an owner of an NFT transfers ownership of the NFT to another person; In other words, it is an act of giving a valuable asset to another person. So this type of interaction conveys a strong connection between two accounts and becomes a subject of interest in detecting suspicious transactions. 

\begin{figure}[!ht]
    \centering
    \includegraphics[width=0.95\textwidth]{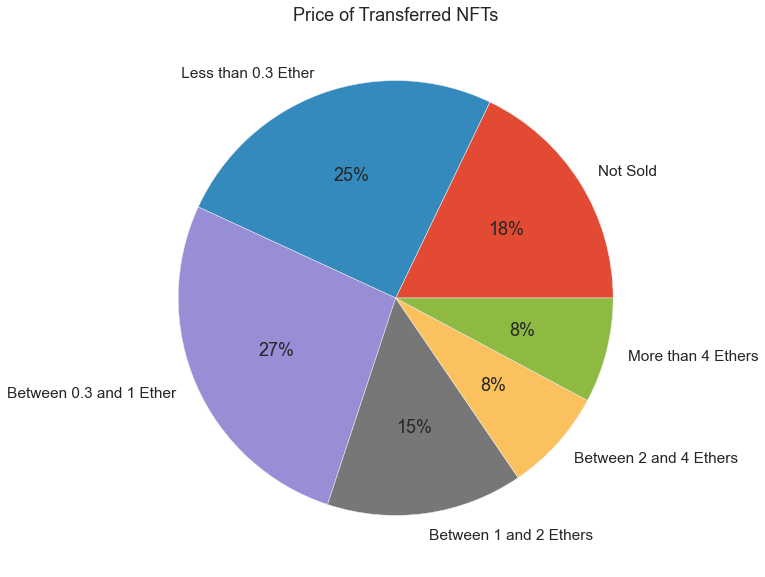}
    \caption{Prices of Transferred NFTs pie chart.}
    \label{fig:transfer_price}
\end{figure}

An interesting fact about NFT transfers that occur on Foundation is that the value of an average transferred NFT is more than the value of an average NFT. As indicated in figure \ref{fig:transfer_price}, a substantial 82 percent of all NFTs that have been transferred have successfully been sold, whereas the same percent for an average NFT on Foundation is only 36.10 percent. This intriguing and significant difference further elevates the need for inspecting the transfers that take place on Foundation.

To investigate further and analyze the NFT transfers that are happening on Foundation, we built a directed graph where each node is an account and the weight of each directed edge between node A, and B indicates the number of NFTs that A has transferred to B. Metrics related to this graph and its largest connected component are provided in table \ref{tab:transfer-graph-metrics}. As it is shown in table \ref{tab:transfer-graph-metrics}, this graph is sparse as it has a very low density, and it consists of 751 separate connected components, and among the connected components of this graph, the largest connected component has 788 nodes while the second-largest connected component has 28 nodes, and also many components consist of only two nodes. 

\begin{figure}[!ht]
    \centering
    \includegraphics[width=0.95\textwidth]{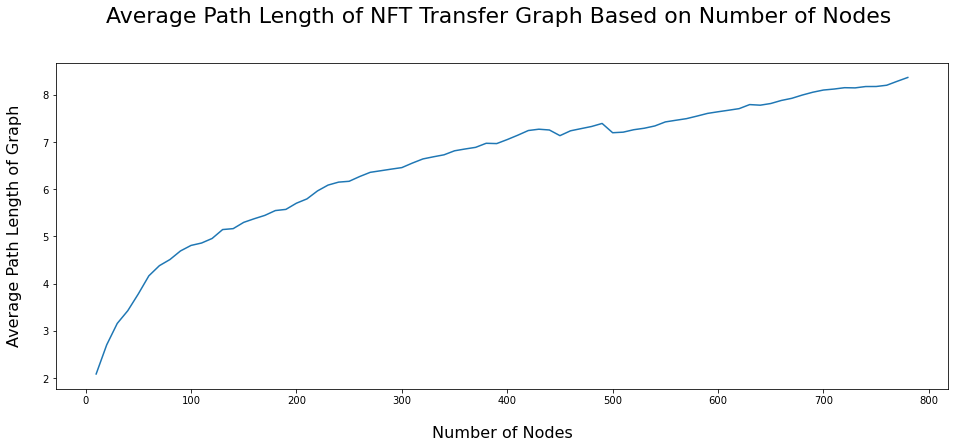}
    \caption{Progression of average path length in largest connected component of NFT transfers graph, based on number of nodes.}
    \label{fig:transfer_shortest_path_by_nodes}
\end{figure}

There are several characteristics in the NFTs transfers graph that suggest this graph is a small-world graph. A graph can be considered to be a small-world graph when it displays two characteristics from itself, the small-world effect and high clustering \cite{watts1998collective}. The small-world effect, which is also present in random graphs, can be observed when two nodes can reach each other within a handful of steps. Quantitatively speaking, the average path length and the radius of a graph with the small-world effect will be relatively low. Also, the average path length of the graph increases fairly slowly as a function of the number of nodes in the graph \cite{albert2002statistical}. By observing Table \ref{tab:transfer-graph-metrics}, we can see that even though this graph has a very low density, its radius and Average Path Length are low. Moreover, In figure \ref{fig:transfer_shortest_path_by_nodes}, Average Path length as a function of the number of nodes has been depicted, and it can be seen that Average Path Length increases rather slowly. 

Clustering in a graph means that when two nodes in a graph have a common neighbor, they will have a tendency to have a connection with each other as well, resulting in more triads in the graph, and this concept can be captured via the clustering coefficient \cite{watts1998collective}. Unlike the small world-effect which is present in both random graphs and small-world graphs, we can only observe high clustering coefficients in small-world graphs, so to measure whether the NFTs' transfer graph that we built exhibits high clustering property or not, we generated 10000 random graphs with the same number of nodes and edges as our own graph using Erdős–Rényi method \cite{erdos1960evolution}, and compared the clustering coefficient of our graph, with the average clustering coefficients of the generated random graphs. As it can be seen from table \ref{tab:transfer-graph-metrics}, the clustering coefficient in the NFT transfer graph was 0.014, while the average clustering coefficient from the generated random graphs was only 0.0002, suggesting that the NFT transfer graph does have the property of having a higher clustering coefficient that the small world graphs display.

Many real-world graphs with distinct subjects have been proven to be small-world graphs, and as we have demonstrated above, the NFT transfers on Foundation graph is also another of these graphs. The fact that the NFT transfers graph is a small-world graph with high clustering coefficient leads to some intriguing observations. The high clustering means this graph has a significant number of highly interrelated nodes that transfer considerable numbers of NFTs, which translate to money within their clusters. These high concentrations of asset transference within certain groups of Ethereum wallets are notable as they may indicate possible money laundering or other malicious actions. Thus, the NFT transfers graph and the highly clustered groups of people in this graph could prove to be crucial points of focus for anomaly detection studies that wish to identify malicious acts.

\begin{table}[]
\centering
\caption{Measures and metrics related to NFT transfers graph and its largest connected component.}
\label{tab:transfer-graph-metrics}
\begin{tabular}{c|c|c|}
\cline{2-3}
                                                                                       & \begin{tabular}[c]{@{}c@{}}NFT\\ Transfer Graph\end{tabular} & \begin{tabular}[c]{@{}c@{}}Largest Connected Component of \\ Transfer Graph\end{tabular} \\ \hline
\multicolumn{1}{|c|}{Nodes}                                                            & 2546                                                         & 788                                                                                      \\ \hline
\multicolumn{1}{|c|}{Edges}                                                            & 2013                                                         & 865                                                                                      \\ \hline
\multicolumn{1}{|c|}{\begin{tabular}[c]{@{}c@{}}Connected\\  Components\end{tabular}}  & 751                                                          & 1                                                                                        \\ \hline
\multicolumn{1}{|c|}{Average Degree}                                                   & 1.581                                                        & 2.195                                                                                    \\ \hline
\multicolumn{1}{|c|}{Maximum Degree}                                                   & 34                                                           & 23                                                                                       \\ \hline
\multicolumn{1}{|c|}{Diameter}                                                         & 23                                                           & 23                                                                                       \\ \hline
\multicolumn{1}{|c|}{\begin{tabular}[c]{@{}c@{}}Average \\ Path Length\end{tabular}}   & 8.411                                                        & 8.411                                                                                    \\ \hline
\multicolumn{1}{|c|}{Density}                                                          & 0.0003                                                       & 0.0014                                                                                   \\ \hline
\multicolumn{1}{|c|}{\begin{tabular}[c]{@{}c@{}}Clustering\\ Coefficient\end{tabular}} & 0.014                                                        & 0.011                                                                                    \\ \hline
\multicolumn{1}{|c|}{\begin{tabular}[c]{@{}c@{}}Degree\\ Assortativity\end{tabular}}   & -0.051                                                       & -0.24                                                                                    \\ \hline
\multicolumn{1}{|c|}{Transitivity}                                                     & 0.011                                                        & 0.008                                                                                    \\ \hline
\end{tabular}
\end{table}

\begin{figure}[!ht]
    \centering
    \includegraphics[width=0.95\textwidth]{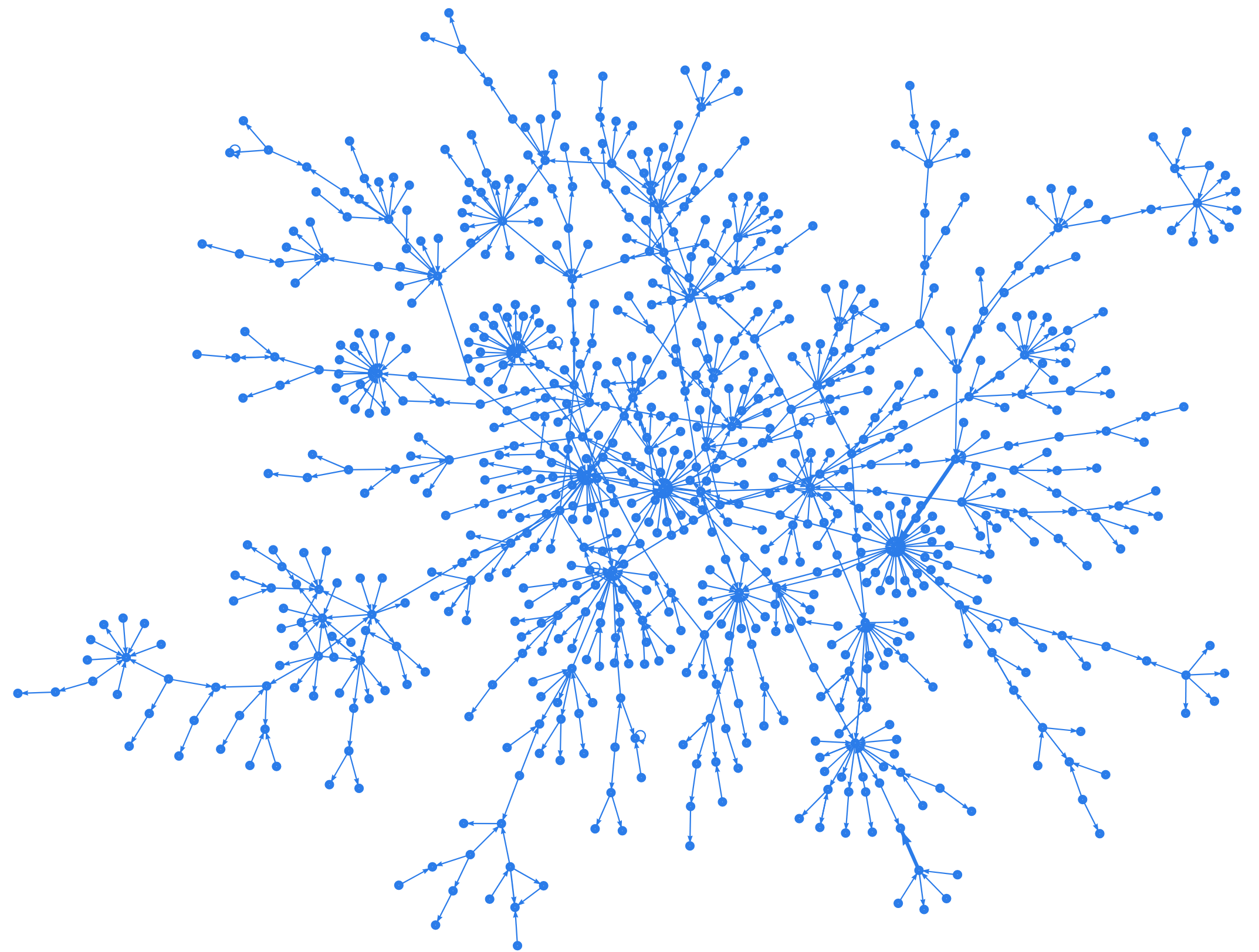}
    \caption{Depiction of largest connected component of NFT transfers network.}
    \label{fig:transferG_0}
\end{figure}

\section{Content-based Price Analysis} \label{s:cbpa}
% \section{Content-based Price Analysis}

\subsection{Approach}

In this chapter, we would like to conduct a price analysis based on the content of NFTs. In particular, we would like to create a model solely based on the raw pixel values on NFT and investigate if there is a meaningful relationship between content and the price. To achieve this, we will incorporate contemporary techniques from the field of Computer Vision and Deep Learning to understand images and extract information about the contents of each NFT.  

One naive approach to utilizing Computer Vision techniques for this problem is to perform object detection on images and categorize these detected objects. This approach has lots of shortcomings, i.g, the list of objects is not exhaustive, and we can not detect all objects (ImageNet dataset \cite{krizhevsky2012imagenet}  has only 1000 categories), more importantly, art images are considered out of distribution samples, hence the hard problem, for pre-trained neural networks. To get away from this issue, we look into building sets of similarity clusters based on the content and will do the price analysis in each of these groups.

\subsection{Clustering based on Similarity Scores}

With the recent advances in Computer Vision and Deep Learning, it is possible to understand images and the story behind an image. Standard image recognition families of neural networks like VGG \cite{simonyan2014very} and ResNet \cite{he2016deep} are capable of detecting objects inside a photograph. These networks have been trained on datasets of natural images, e.g., ImageNet, and out of the box, they are not suitable for artificially generated photos. Several attempts have been made to improve these networks' generalization on out-of-distribution samples and even on paintings and art images  \cite{risi2102.06529}. That being said, current models can be used as a deep feature extractor to generate a dense representation of an image's content.

We would like to create a group of NFTs based on a similarity measurement, i.e., put visually similar NFTs in one group. To achieve this, we will calculate cosine distance as a similarity metric between two images. While it is possible to do cosine similarity on the pixel space, the resulting number is not meaningful and extremely sensitive to the object's location on the image. Instead, we will use image embedding generated by a neural network. For this purpose, we used a pre-trained ResNet101 model (downloaded from PyTorch \cite{paszke2019pytorch} model hub) to extract embedding vectors for images in our dataset. We stored the output of \texttt{layer\_4} in the ResNet101 model which has the volume of \texttt{2048x7x7}. After flattening this feature volume, we will get a vector of size 100,352. The aforementioned process will transform each image of any size into a vector of size 100,352. The similarity score between two images is the result of the cosine similarity function on correspondence vectors for those images.

For constructing similarity groups, we found the top 5 nearest images to a given query image. Among all NFTs we have collected, only 23,470 of them were images (with various formats), and 7,634 of these had at least one successful auction. Since we want to price analyzing, we do this on the NFTs with at least one successful auction. That’s it, for every NFTs that had successfully sold at auction, we extracted the top 5 similar NFTs that also were sold on an auction. Since we are dealing with high-dimensional data and we need to perform multiple queries, we need a specialized data structure to have a quick query processing time. To address this, we used the FAISS library \cite{JDH17} to build an index on top of embeddings based on the cosine distance of the elements. After building such an indexing system, it is easy to find similar images for an arbitrary query image.

\begin{figure}[!ht]
\centering
\includegraphics[width=0.95\linewidth]{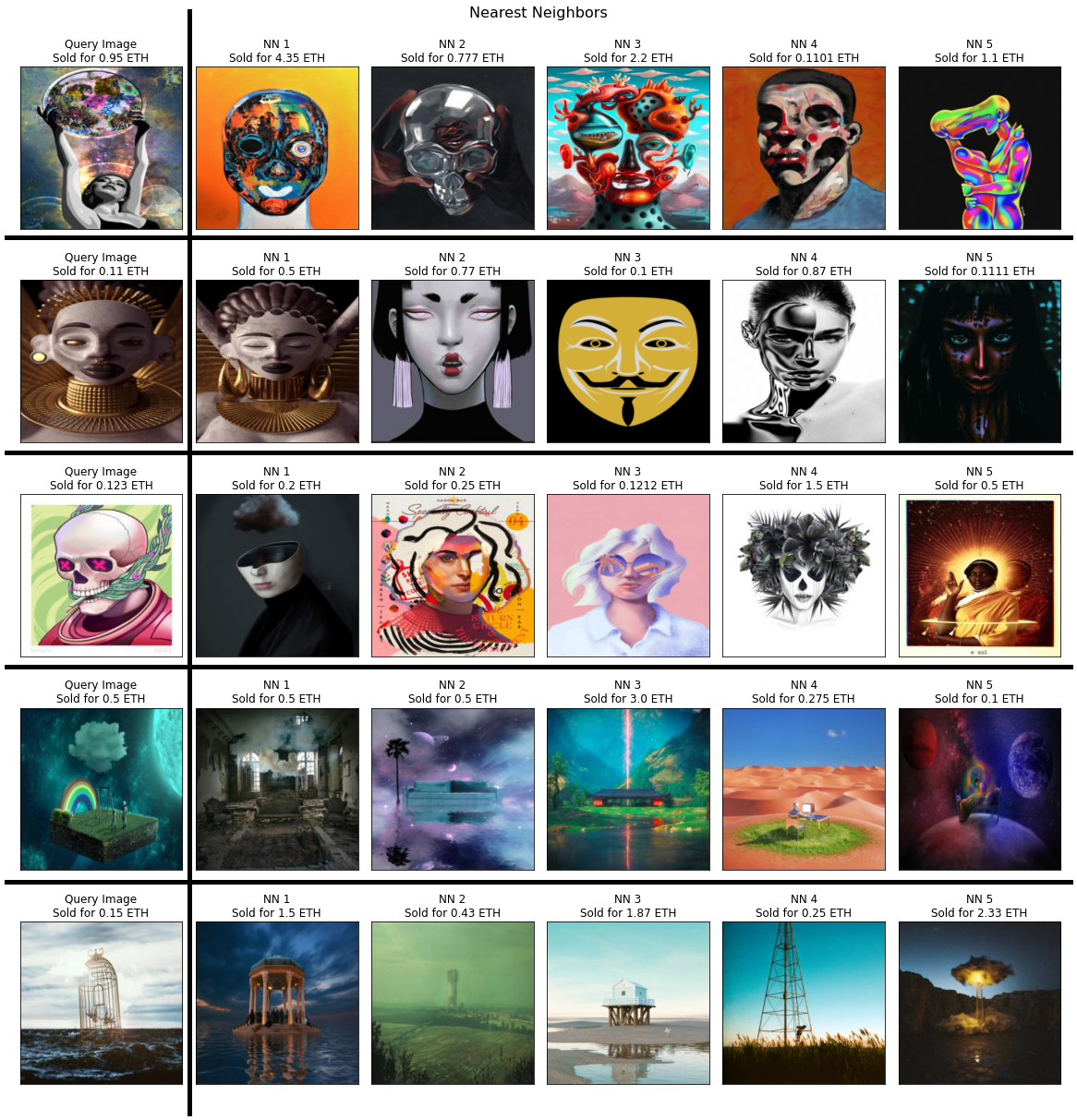}
\caption{Nearest Neighbors based on Embeddings}
\label{fig:nns}
\end{figure}

In Figure \ref{fig:nns}, Each row shows an NFT along with its five nearest neighbors. As we can see, there exists some understandable pattern along each row. For instance, The first row contains images of a human head, the second and the third row includes images of landscapes, etc. This suggests that embedding generated by ResNet101 has a high degree of discriminative power to distinguish different images and can be used as a good similarity proxy for our task.

\subsection{Price Analysis}

\begin{figure}[!ht]
\centering
\includegraphics[width=0.95\linewidth]{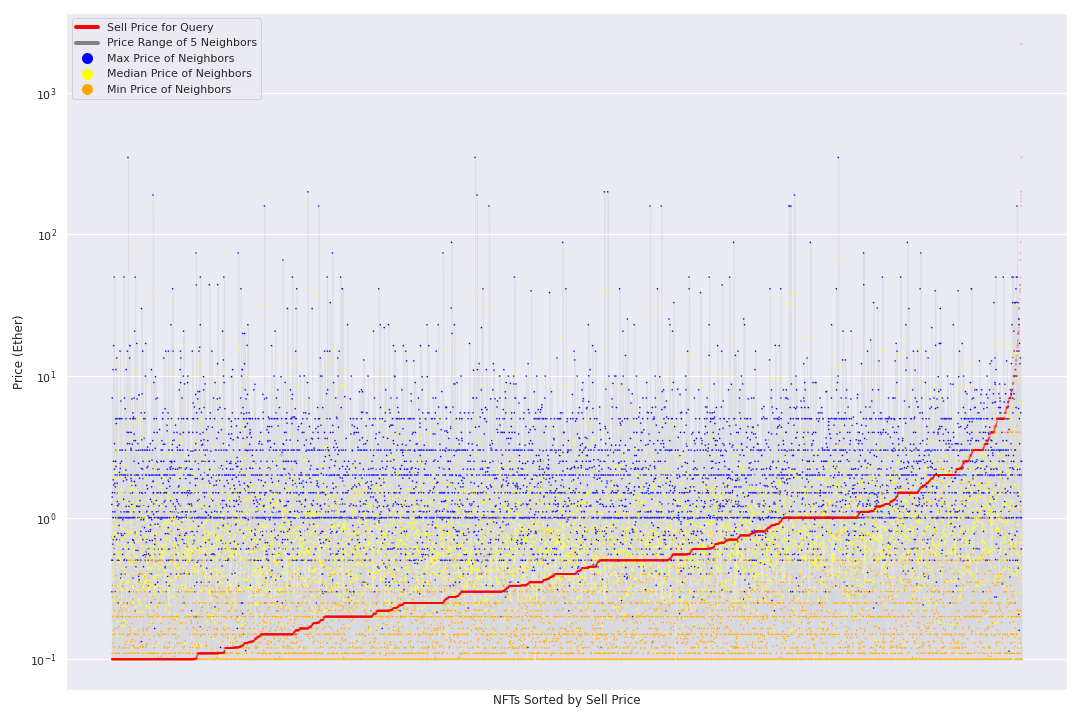}
\caption{Price variation for a given NFT and its neighbors}
\label{fig:nn_dists}
\end{figure}

In the last section, we introduced an approach to group NFTs based on their content similarity. For each query image, we created a group containing five other NFTs with the highest similarity score to the given query. In this stage, we want to analyze price variation inside each of these similarity groups. In the figure \ref{fig:nn_dists} we plotted the price of all sold NFTs from the previous step along with the price of their nearest neighbors. On the X-axis, we show a sorted list of NFTs (red graph), and each dot along the Y-axis indicates the price for its neighbors. This chart shows extreme price variation on each step, regardless of the selling price for the center NFT.

\begin{figure}[!ht]
\centering
\includegraphics[width=0.95\linewidth]{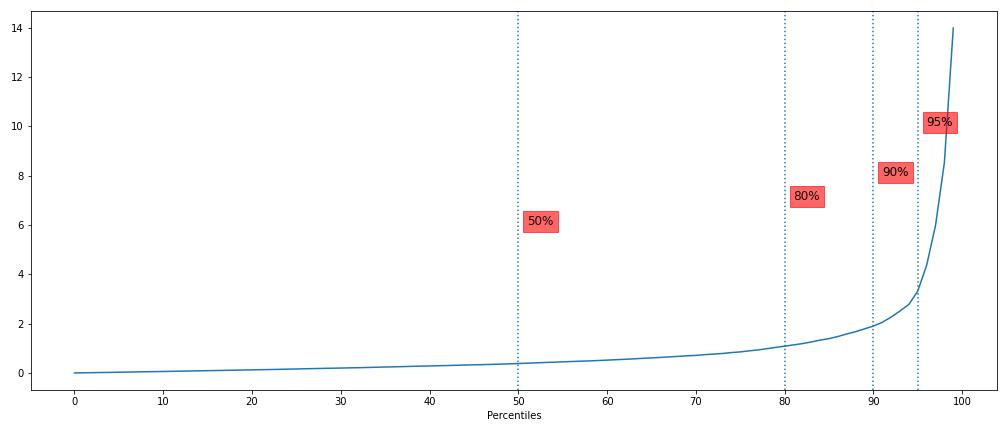}
\caption{Distance between selling price and average selling price for neighbors}
\label{fig:nn_percentiles}
\end{figure}

To quantify this price variation, for every query point in our dataset, we calculated the distance between the average selling price for the neighbors and the selling price for the query. The average price for neighbors gives us an estimate of how similar artworks perform at an auction. After calculating the aforementioned distance, we found that the price of 80\% of all sold NFTs falls within a 1 Ether range from its neighbors. 80\% of the time, the selling price for artwork is less than 1 Ether from the average selling price for the top 5 similar artworks. Figure \ref{fig:nn_percentiles} shows a distribution plot for the distance between the selling price and average price for similar NFTs.

To show that similarity groups play a role here, we also calculated the distance between the price of each query point and the average of all sold NFTs. Only 46\% of the time, the spread between these two prices are less than 1 Ether, which suggests that prices on each similarity group are more coherent.

\section{Conclusion} \label{s:conclusion}
% \section{Conclusion}

Non-Fungible Tokens have constantly been gaining ground within the crypto community, as people have realized the potentials that lie in NFTs. In this study, we first crawled the data of 65 thousand NFT auctions that had occurred on the Foundation platform. We then explored auction activities and some underlying dynamics on Foundation, such as the potential for speculative behaviors for selling second-hand arts. We found that while 36.10\% of NFTs were sold on their first auctions, only 3.97\% of NFTs that were listed for a second auction were sold again, which suggests that speculative behavior on Foundation may not be prevalent yet. 

Moreover, our study explores the possibility of undisclosed agreements to exchange creator account invitation letters on Foundation. Through analyzing the auctions in which a user had bought an NFT from the same user who had invited them, we estimate that the value of a creator account on Foundation is approximately 0.3716 Ether.

We then analyzed NFT transfers that occur on Foundation. Based on network analysis metrics, Our observation shows that the NFT transfers network exhibits high clustering and small-world graph characteristics. There are groups of highly interrelated users in the NFT transfers graph with high rates of transferring NFTs among themselves, and these clusters can become valuable points of focus for future studies about anomaly detection in NFT transactions.

At last, we performed a price analysis solely based on the content of artworks, i.e., based on the pixels of the NFTs with image formats. We showed that there exists a meaningful relationship among the prices of NFTs, which have a high similarity with each other. Moreover, we showed that for 80\% of all sold NFTs, their selling price lies in 1 Ether distance from the average price for the five most similar NFTs. This suggests that we can build a recommendation system based on the content of artworks, that firstly, finds similar artworks, and secondly, estimates a starting price for a new minted NFT.

\section*{Code and Data Availability Statement} \label{s:data}
Implementation of all the prediction methods and the used data is available on
\url{https://github.com/taesiri/NFTFoundation}

\section*{Acknowledgements}
The authors did not receive support from any organization for the submitted work.

\section*{Conflict Of Interest}
No funds, grants, or other support was received. All authors certify that they have no affiliations with or involvement in any organization or entity with any financial interest or non-financial interest in the subject matter or materials discussed in this manuscript.

\bibliographystyle{abbrvnat}
\bibliography{main}  %%% Uncomment this line and comment out the ``thebibliography'' section below to use the external .bib file (using bibtex) .





% \end{thebibliography}

\end{document}